\documentclass[aps,prl,twocolumn,floatfix,amsmath,amssymb,showpacs,
               superscriptaddress]{revtex4-1}

\usepackage{graphicx,placeins}
\usepackage{float}
\usepackage{color}
\usepackage{dcolumn}
\usepackage{bm}
\usepackage{epsfig,psfrag,amsmath,amssymb,float}
\input{epsf}
\usepackage{hyperref}
\usepackage[percent]{overpic}
\hypersetup{
    colorlinks=true,
    linkcolor=blue,
    citecolor=blue,
    filecolor=magenta,      
    urlcolor=cyan,
}
\usepackage{array}
\newcolumntype{C}[1]{>{\centering\arraybackslash$}p{#1}<{$}}
\usepackage{titlesec}

\setcitestyle{square}

\bibpunct{[}{]}{,}{n}{}{}
\begin{document}
\title{Kanamori-Moir\'e-Hubbard model for transition metal dichalcogenide homobilayers}

\author{Nitin Kaushal}
\affiliation{Materials Science and Technology Division, Oak Ridge National 
Laboratory, Oak Ridge, Tennessee 37831, USA}




\author{Elbio Dagotto}
\affiliation{Materials Science and Technology Division, Oak Ridge National 
Laboratory, Oak Ridge, Tennessee 37831, USA}
\affiliation{Department of Physics and Astronomy, The University of 
Tennessee, Knoxville, Tennessee 37996, USA}

\date{\today}

\begin{abstract}
{\it{Ab-initio}} and continuum model studies predicted that the $\Gamma$ valley transition metal dichalcogenide (TMD) homobilayers could simulate the conventional multi-orbital Hubbard model on the moir\'e Honeycomb lattice. Here, we perform the Wannierization starting from the continuum model and show that a more general Kanamori-Moir\'e-Hubbard model emerges, beyond the extensively studied standard multi-orbital Hubbard model, which can be used to investigate the many-body physics in the $\Gamma$ valley TMD homobilayers. Using the unrestricted Hartree-Fock and Lanczos techniques, we study these half-filled multi-orbital moir\'e bands. By constructing the phase diagrams we predict the presence of an antiferromagnetic state and in addition we found unexepected and dominant states, such as a $S=1$ ferromagnetic insulator and a charge density wave state. Our theoretical predictions made using this model can be tested in future experiments on the $\Gamma$ valley TMD homobilayers. 
\end{abstract}
\maketitle

{\it{Introduction.---}}
Transition metal dichalgenide (TMD) moir\'e materials provide unprecedented platforms to study the effect of electronic correlations on flat moir\'e bands~\cite{Mak01,Andrei01,CZhang01,YPan01,McGilly01}. A variety of low-energy Hamiltonians can be realized in these TMD moir\'e materials~\cite{Kennes01}. For example, the $\mathrm{WSe_{2}/WS_{2}}$ heterobilayer simulates the one-orbital triangular lattice Hubbard model~\cite{Tang01,Regan01,CJin02,FWu01}, while the AB-stacked $\mathrm{MoTe_{2}/WSe_{2}}$ leads to non-trivial moir\'e bands demonstrating quantum anomalous Hall effect~\cite{TLi01}. In addition, recent {\it ab-initio} and continuum model calculations have shown that twisted $\Gamma$-valley homobilayers, such as $\mathrm{MoS_{2}}$, $\mathrm{MoSe_{2}}$, and $\mathrm{WS_{2}}$, produce two valence moir\'e bands with Dirac cone mimicking a honeycomb lattice, while the next set of lower energy four moir\'e bands simulates the two-orbital asymmetric $p_{x}$-$p_{y}$ honeycomb lattice model~\cite{Angeli01,Xian01,Vitale01,Wu01}. Moreover, surprisingly in recent ARPES experiments $\Gamma$-valley moir\'e bands have been observed in the twisted $\mathrm{WSe_{2}}$~\cite{Pei01,Gatti01}, rendering it also a candidate material to realize the two-orbital honeycomb lattice model. These findings opens up an exciting avenue to simulate multi-orbital Hubbard-like models in TMD moir\'e materials.

The Kanamori-Hubbard (KH) model~\cite{Kanamori01,DagottoReview} has been extensively studied for many conventional materials where multiple orbitals are active, as in iron based superconductors, iridates, manganites. etc.~\cite{Herbrych01,Herbrych02,Kaushal02,Kim01,DagottoManganites}. The moir\'e potential is shallower than the ionic potential present in conventional materials, leading to relatively broader Wannier functions in moir\'e materials and making non-local correlations important~\cite{Duran01}, which are typically ignored in the often used KH model. This suggests that the theoretical studies of twisted $\Gamma$-valley homobilayers require a model going beyond the standard KH model. In this publication, for the first time we provide a Kanamori-Moir\'e-Hubbard (KMH) model which includes non-local correlations, where the interaction parameters are calculated using the well-localized and accurate Wannier functions of the twisted $\mathrm{MoS_{2}}$ bilayer~\cite{Naik01,Naik02,Liao01}. The importance of the KHM model is depicted by discussing the effective dielectric constant $\epsilon$ vs the twist angle $\theta$ phase diagrams for the half-filled KMH model, unveiling surprising results which definitely cannot be captured by the standard KH model. It is interesting to note that the relevance of the non-local correlations in the flat moir\'e bands of twisted bilayer graphene (TBG) has also been discussed~\cite{Yuan01,Kishino01,Kang01}, so we believe the KMH model can also be used for TBG, but only near magic angles~\cite{Bistritzer01,Cao01,Cao02,Andrei02} unlike in TMD bilayers where the flat bands are present in a larger range of twist angles.

 \begin{figure*}[!ht]
\hspace*{-0.5cm}
\vspace*{0cm}
\begin{overpic}[width=1.8\columnwidth]{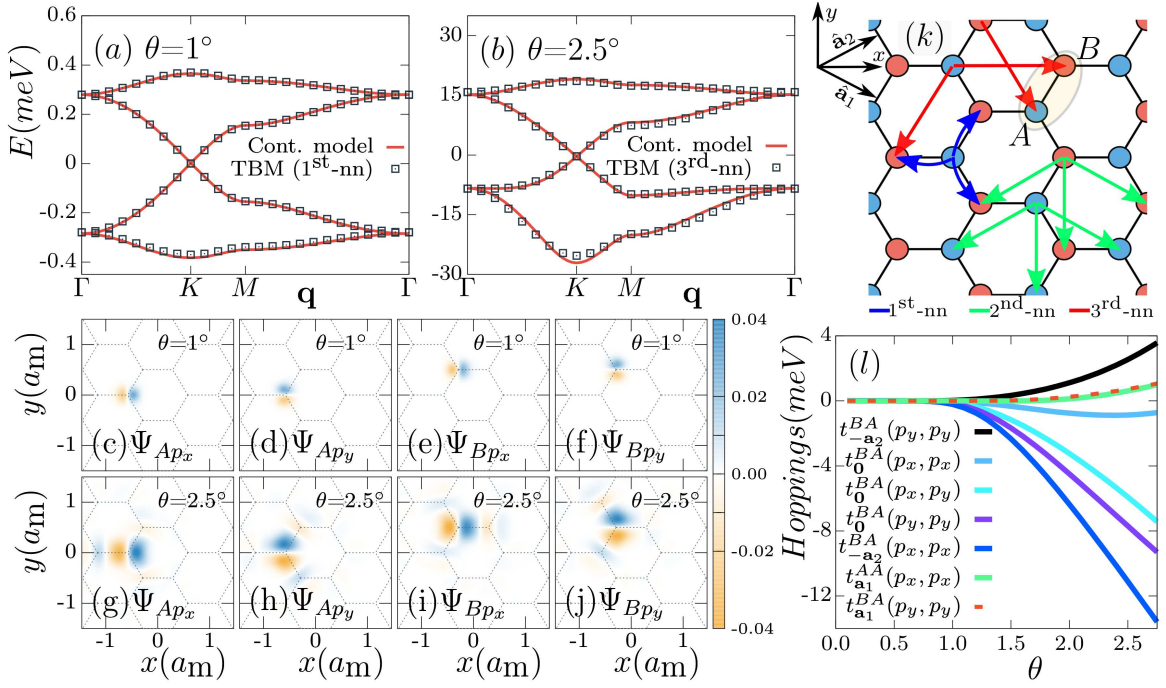}
\end{overpic}
\caption{(a,b) Comparison between  independently calculated band structures using the continuum model and the tight-binding model (TBM), for the twist angles (a) $\theta=1^{\circ}$ and (b) $\theta=2.5^{\circ}$. Wannier functions, calculated using the continuum model Bloch wavefunctions, for twist angles $\theta=1^{\circ}$ and $\theta=2.5^{\circ}$ are shown in panels (c-f) and (g-j), respectively. (k) The honeycomb lattice geometry used in the tight-binding model. The blue, green, and red arrows depicts the nearest, 2nd-nearest, and 3rd-nearest neighbours hoppings, respectively. (l) Evolution of the dominant hopping parameters with the twist angle.}
\label{fig1}
\end{figure*}

{\it{Wannierization and tight-binding model.---}} We calculate the  moir\'e bands structure and the Bloch states using the continuum moir\'e Hamiltonian $H=-\hbar k^{2}/2m^{*} + \Delta({\bf{r}})$. The moir\'e potential $\Delta(\bf{r})$ is defined as $\Delta({\bf{r}})=\sum_{s}\sum_{j=1}^{6}V_{s}e^{i({\mathbf{g}}_{j}^{s}.{\mathbf{r}} + \phi_{s})}$, where ${\bf g}_{j}^{s}$ are the moir\'e reciprocal lattice vectors connecting to $s$-th nearest neighbour. The model parameters $\{ V_{s}, \phi_{s}\}$ are fixed following earlier studies \cite{Angeli01}, considering the $\mathrm{MoS_{2}}$ homobilayer, so that the band structure obtained from continuum model and {\it ab-initio} matches very well. All of our predictions are also valid for other $\Gamma$ valley homobilayers like $\mathrm{MoSe_{2}}$ and $\mathrm{WS_{2}}$. The valence bands closest to the chemical potential can be described by a one-orbital tight-binding model on a honeycomb lattice, see~\cite{Angeli01,Xian01}. Here, we focus on the second-set of 4 composite valence bands, which can be described by a two-orbital $p_{x}$-$p_{y}$ tight-binding model on the honeycomb lattice. Until now, the Wannier functions have not been calculated for these set of bands. We perform Wannierization, using projection technique~\cite{Marzari01,Marzari02}, to obtain 4 well-localized Wannier functions, two on each sublattice namely $A$ and $B$, see Fig.~\ref{fig1}(k) (for details see supplementary~\cite{supple}). The calculated Wannier fuctions have nodes at the moir\'e sites and a pair of lobes like in the $p$-orbitals of the hydrogen atom, as shown in Fig.~\ref{fig1}(c-f) and Fig.~\ref{fig1}(g-j) for twist angles $1^{\circ}$ and $2.5^{\circ}$, respectively. We noticed that $\Psi_{A(B)p_{x}}({\bf{r}})$ cannot be obtained by a $90^{\circ}$ rotation of $\Psi_{A(B)p_{y}}({\bf{r}})$ unlike in the ideal $p_{x}$-$p_{y}$ orbitals, which follows from the absence of full rotational symmetry in the moir\'e potential. Moreover, we found $\Psi_{Bp_{x(y)}}({\bf{r}})=-\Psi_{Ap_{x(y)}}(-{\bf r})$ due the inversion symmetry of the moir\'e potential on two sublattices given by $\Delta({\bf r}-{\bf R}_{A})$=$\Delta(-{\bf r}-{\bf R}_{B})$.

Using the above Wannier functions, we calculated the hopping parameters for the two-orbital tight-binding model on the honeycomb lattice, up to third nearest-neighbour using $t_{{\bf j}-{\bf i}}^{SS^{'}}(\mu,\nu)=\langle\Psi_{S\mu}^{\bf j}|H|\Psi_{S^{'}\nu}^{\bf i}\rangle$ (for details see Supplementary Material~\cite{supple}), where the $\{{\bf i}, {\bf j}\}$, $\{S,S^{'}\}$, and $\{\mu, \nu\}$ indices denotes unit-cell, sublattice,  and orbitals ($p_{x}$ or $p_{y}$), respectively. We write the kinetic energy as $H_{\textrm{K.E.}}=\sum_{\bf{i}\sigma}K_{\bf{i}\sigma}^{1} + K_{\bf{i}\sigma}^{2} + K_{\bf{i}\sigma}^{3}$, where the terms $K_{\bf{i}\sigma}^{n}$ consists of hoppings between the $n^{th}$ nearest neighbour sites in the honeycomb lattice. The hopping connections up to the 3rd nearest-neighbour are pictorially shown in Fig.~\ref{fig1}(k). $K_{\bf{i}\sigma}^{1}$ is presented below: 
\begin{equation}\label{1stNNhop}
K_{\bf{i}\sigma}^{1}=\sum_{\substack{\nu,\mu\in\{p_{x},p_{y}\} \\ \bf{r}\in \{0,-\bf{a}_{2},\bf{a}_{1}-\bf{a}_{2}\}}}t_{\bf{r}}^{BA}(\mu,\nu)c_{{\bf{i}+\bf{r}}B_{\mu}\sigma}^{\dagger}c_{{\bf{i}}A_{\nu}\sigma} + h.c.
\end{equation}

The $K_{\bf{i}\sigma}^{2(3)}$ terms can be written similarly, as shown in supplementary~\cite{supple}. The 1st 
nearest-neighbour hopping term $K_{\bf{i}\sigma}^{1}$, shown in eq. \ref{1stNNhop}, depends on three 2$\times$2 matrices namely $\{t_{\bf{0}}^{BA}, t_{-\bf{a}_{2}}^{BA}$, $t_{\bf{a}_{1}-\bf{a}_{2}}^{BA}\}$. Similarly six 2$\times$2 matrices $\{t_{\bf{a}_{1}}^{AA}, t_{-\bf{a}_{2}}^{AA}, t_{\bf{a}_{1}-\bf{a}_{2}}^{AA}, t_{\bf{a}_{1}}^{BB}, t_{-\bf{a}_{2}}^{BB}, t_{\bf{a}_{1}-\bf{a}_{2}}^{BB} \}$ and three 2$\times$2 matrices $\{ t_{\bf{a}_{1}}^{AB}, t_{\bf{a}_{1}}^{BA}, t_{\bf{a}_{1}-2\bf{a}_{2}}^{BA} \}$ are required for the 2nd and 3rd nearest-neighbour hoppings, respectively. All of these 12 matrices are dependent on $\theta$. We found a good match between the band-structure calculated using the above tight-binding model and the continuum model, as shown in Fig.~\ref{fig1}(a,b), suggesting that we have accurate Wannier functions. We noticed that for $\theta \lessapprox1.2$, only nearest-neighbour hoppings are enough to obtain the correct band-structure, as shown in Fig.~\ref{fig1}(a) for $\theta=1.0$. However, for larger $\theta$ longer-range hoppings are required to reproduce the continuum model results (see Fig.~\ref{fig1}(b), for $\theta=2.5$). We show the evolution of the some dominant hopping parameters in Fig.~\ref{fig1}(l), depicting the exponential fast growth of hoppings with $\theta$.

{\it{Interaction parameters and Kanamori-Moir\'e-Hubbard model.---}} Now we will derive the Coulomb interaction between the fermions in the Wannier states discussed above. The generic interaction term can be written as:
\begin{equation}
H_{\textrm{Int}}=1/2\sum_{\substack{\mathbf{i},\mathbf{j},\mathbf{k},\mathbf{l},\\ \alpha,\beta,\gamma,\delta,\\ \sigma, \sigma{'} }}V_{\mathbf{ijkl}}^{\alpha\beta\gamma\delta}c_{\bf{i}\alpha\sigma}^{\dagger}c_{\bf{j}\beta\sigma{'}}^{\dagger}c_{\bf{l}\delta\sigma{'}}c_{\bf{k}\gamma\sigma},
\label{Int_VTerm}
\end{equation}
where $V_{\mathbf{ijkl}}^{\alpha\beta\gamma\delta}=\langle \Psi^{\bf{i}}_{\alpha}\Psi^{\bf{j}}_{\beta}|V|\Psi^{\bf{k}}_{\gamma}\Psi^{\bf{l}}_{\delta} \rangle$ and $V=e^{2}/\epsilon|{\bf r}_{1}-{\bf r}_{2}|$. $\epsilon$ is produced by the surrounding dielectric enviroment, such as nearby h-BN layers. The exact value of $\epsilon$ is not known so we keep it as a free parameter. The $\{\alpha,\beta,\gamma,\delta\}$ indices represent the sublattice $S$ and the orbital $\mu$ via $\alpha=2S+\mu=S_{\mu}$, where the sublattice $A(B)=0(1)$ and the orbital $p_{x}(p_{y})=0(1)$.

In the present work, for simplicity, we limit the non-local Coulomb interactions only up to nearest-neighbour sites of the honeycomb lattice. A priori, the longer range interactions are not expected to be very relevant at and near half-filling~\cite{Pan01}. To study Wigner crystals at fractional fillings, the approximate longer range interactions can be easily included by assuming the $(\frac{1}{|r|}-\frac{1}{\sqrt{r^{2}+d^2}})$ functional form, where $d$ is the screening length~\cite{Kaushal01,Pan02,Duran02}. The Coulomb interaction term which includes up to nearest-neighbour interactions can be divided into three parts, $H_{\textrm{Int}} = \sum_{\bf{i}} H_{\bf{i}} + H_{\bf{i},{\bf{i}\textrm{-}}{\bf{a}_{2}}}^{AB} + H_{\bf{i},{\bf{i}\textrm{+}\bf{a}_{1}}{\textrm{-}}{\bf{a}_{2}}}^{AB}$, where $\bf{i}$ is the unit cell index and $\bf{a_{(1,2)}}$ are the Bravais lattice vectors. The first part $H_{\bf{i}}$ consists of all the Coulomb interactions possible within the unit cell $\bf{i}$, including both local and nearest-neighbour interactions given by $V_{{\bf{iiii}}}^{\alpha\beta\gamma\delta}$ (total $4^4$ terms). The second $H_{\bf{i},{\bf{i}\textrm{-}}{\bf{a}_{2}}}^{AB}$ and third parts $H_{\bf{i},{\bf{i}\textrm{+}\bf{a}_{1}}{\textrm{-}}{\bf{a}_{2}}}^{AB}$ contains the Coulomb interactions between the nearest-neighbour sites belonging to different unit cells. Now we will discuss the $H_{\bf{i}}$ term in detail; the other two terms are very similar and shown in the supplementary. $H_{\bf{i}}$  is shown in Eq.~\ref{KMH}, where ${\bf{S}}_{{\bf{i}}\alpha}$=$\frac{1}{2}\sum_{s,s^{'}} c_{{\bf i}\alpha s}^{\dagger} {\bm{\tau}}_{ss^{'}} c_{{\bf i}\alpha s^{'}}$ represent the spin at unit cell ${\bf i}$, orbital $\mu$=mod$(\alpha,2)$, and sublattice=$(\alpha-\mu)/2$. The pair anhilation operator is defined as $P_{\bf{i}\alpha}=c_{\bf{i}\alpha\downarrow}c_{\bf{i}\alpha\uparrow}$. $s=1(-1)$ for $\sigma=\uparrow(\downarrow)$, and the set $\mathbb{S}=\{ \{0123\},\{0132\}, \{0213\}\}$. 

\begin{multline}
H_{\bf{i}}  = U_{0}\sum_{\alpha}n_{\bf{i}\alpha\uparrow}n_{\bf{i}\alpha\downarrow}
+\sum_{\alpha<\beta}(U_{\alpha\beta}-\frac{J_{\alpha\beta}}{2})n_{\bf{i}\alpha}n_{\bf{i}\beta}
\\
-2\sum_{\alpha<\beta}J_{\alpha\beta}{\bf S}_{\bf{i}\alpha}\cdot{\bf S}_{\bf{i}\beta} 
+ \sum_{\alpha<\beta}J_{\alpha\beta}(P_{\bf{i}\alpha}^{\dagger}P_{\bf{i}\beta} + h.c.)
\\
+1/2\sum_{\substack{\sigma,\sigma{'}, \alpha\ne\beta\ne\gamma}} (A_{\beta\alpha\gamma}-\delta_{\sigma\sigma{'}}\tilde{J}_{\beta\alpha\gamma})(c_{\bf{i}\alpha\sigma}^{\dagger}c_{\bf{i}\gamma\sigma}n_{\bf{i}\beta,\sigma{'}} + h.c.)
\\
+\sum_{\sigma, \alpha\ne\beta} \tilde{A}_{\alpha\beta}(c_{\bf{i}\alpha\sigma}^{\dagger}c_{\bf{i}\beta\sigma}n_{\bf{i}\beta \bar{\sigma}} + h.c.) 
\\
-\sum_{\alpha\ne\beta\ne\gamma}\tilde{J}_{\alpha\gamma\beta}(S_{\bf{i}\alpha}^{+}c_{\bf{i}\beta\downarrow}^{\dagger}c_{\bf{i}\gamma\uparrow} + h.c.)
\\ 
+1/2\sum_{\substack{\sigma,\alpha\ne\beta\ne\gamma}}\tilde{J}_{\alpha\gamma\beta}s(P_{\bf{i}\alpha}^{\dagger}c_{\bf{i}\gamma\bar{\sigma}}c_{\bf{i}\beta\sigma} + h.c.)
\\
+\sum_{\substack{\sigma,\sigma{'}, \\ \{\alpha\beta\gamma\delta\}\in {\mathbb{S}} }}T_{\alpha\beta\gamma\delta}(c_{\bf{i}\alpha\sigma}^{\dagger}c_{\bf{i}\gamma\sigma} ( c_{\bf{i}\beta\sigma{'}}^{\dagger}c_{\bf{i}\delta\sigma{'}} + c_{\bf{i}\delta\sigma{'}}^{\dagger}c_{\bf{i}\beta\sigma{'}})+h.c.)
\label{KMH}
\end{multline}

Equation~\ref{KMH} encompasses all $4^4$ intra-unit cell interaction terms. The first four terms look similar to the conventional multiorbital Hubbard model, but here they capture the non-local interactions as well. This is the first time such a model is shown.
\begin{figure}[!h]
\hspace*{-0.5cm}
\vspace*{0cm}
\begin{overpic}[width=0.9\columnwidth]{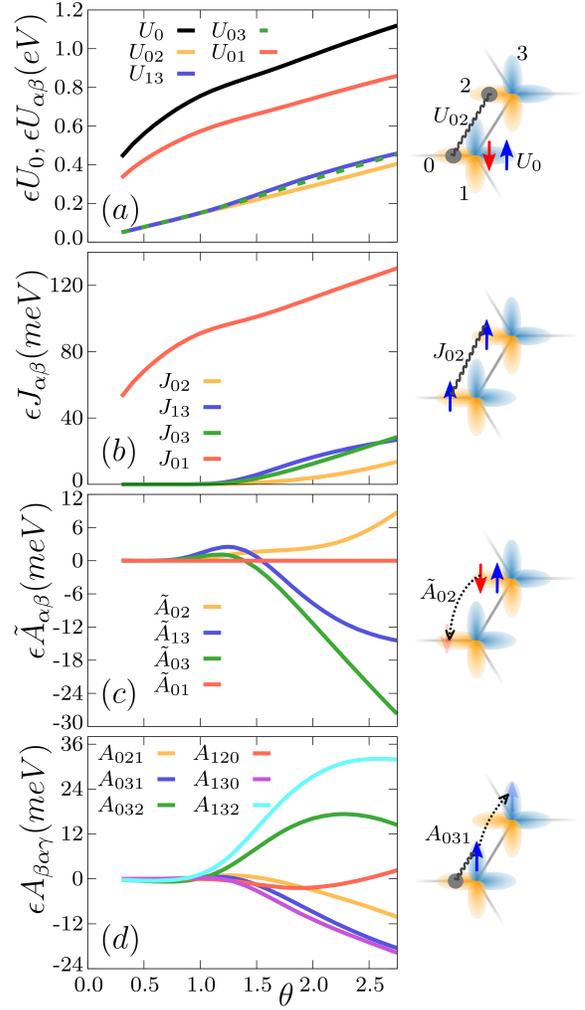}
\end{overpic}
\caption{(a) The onsite intra-orbital Hubbard repulsion $U_{0}$, onsite inter-orbital Hubbard repulsion $U_{01}$, and the orbital resolved nearest-neighbour Hubbard repulsion parameters $\{U_{02},U_{03},U_{13}\}$ shown for various twist angle $\theta$ values. (b) The onsite Hund's coupling $J_{01}$ and the orbital resolved nearest neighbour direct-exchange parameters $\{J_{02},J_{03},J_{13}\}$ as a function of $\theta$. (c,d) Evolution of nearest-neighbour interaction assisted hoppings with $\theta$, requiring electron pair in the same-orbital (c) or on the same-site (d). $\epsilon$ is the effective dielectric constant.   
 }
\label{fig2}
\end{figure}

The first term is the standard onsite intra-orbital Hubbard repulsion, where $U_{0}=V_{\bf{iiii}}^{\alpha\alpha\alpha\alpha}$ (same for all $\alpha$'s). The second term incorporates the onsite inter-orbital density-density repulsions via parameters \{$U_{01}$, $U_{23}$, $J_{01}$, $J_{23}$\} and the non-local orbital resolved repulsions via parameters like $U_{02}$, $J_{02}$, etc., where $U_{\alpha\beta}=V_{\bf{iiii}}^{\alpha\beta\alpha\beta}$ and $J_{\alpha\beta}=V_{\bf{iiii}}^{\alpha\beta\beta\alpha}$. The well known local Hund's coupling is present in the third term via the dominant $J_{01}$ and $J_{23}$ parameters; this term also includes the non-local ferromagnetic direct exchange terms ($J_{02}$,$J_{13}$, $J_{03}$, $J_{12}$). The fourth term incorporate the onsite inter-orbital and non-local pair hopping terms. We also found interaction assisted hoppings (term-5 and term-6), spin-flip hopping accompanied with local spin flip (term-7), and scattering of doublon to different states (term-8) quantified by ($A_{\beta\alpha\gamma}$, $\tilde{A}_{\alpha\beta}$, $\tilde{J}_{\beta\alpha\gamma}$). The remaining interactions are present in term-9.


We show the interaction parameters of first 6 terms as a function of $\theta$ in Fig.~\ref{fig2}. The density-density terms are dominant interactions, see Fig.~\ref{fig2}(a). The onsite intraorbital repulsion ($U_{0}$) suggests that $\epsilon U_{0}/W$ can be of order of 10 to 1000 in real materials, depending on $\theta$, where $W$ is the non-interacting bandwidth. For example, $U_{0}/W$ is about 1200$\epsilon^{-1}$ and 25$\epsilon^{-1}$ for $\theta$=$1^{\circ}$ and $\theta$=$2.5^{\circ}$, respectively. The local Hund's coupling and the non-local ferromagnetic direct exchange is shown in Fig.~\ref{fig2}(b). Fig~\ref{fig2}(c,d) displays the interaction assisted hoppings vs. $\theta$. The rest of the interaction parameters are relatively smaller, and shown in the supplementary. 
We call the total Hamiltonian $H=H_{\textrm{K.E.}} + H_{\textrm{Int}}$ the Kanamori-Moir\'e-Hubbard (KMH) model because of the presence of non-local interaction terms, which are ignored in the standard KH model. These non-local correlations can lead to unexpected results, as shown in the next sextion. It should be noted that the KMH model shown here has larger scope and can be also used for magic-angle TBG and future moir\'e materials addressing multiorbital physics on honeycomb lattice (only the values of hopping and interaction parameters will depend on the specific material).


\begin{figure}[!t]
\hspace*{-0.5cm}
\vspace*{0cm}
\begin{overpic}[width=0.9\columnwidth]{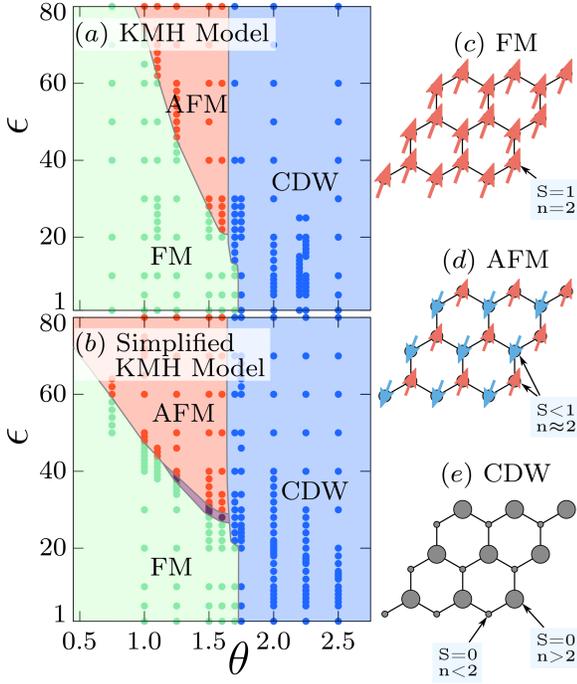}
\end{overpic}
\caption{(a,b) Effective dielectric constant $\epsilon$  vs twist angle $\theta$ phase diagrams for (a) the full Kanamori-Moir\'e-Hubbard (KMH) model and (b) the simplified KMH model, both constructed via unrestricted Hartree-Fock. Panels (c), (d), and (e) show the pictorial representation of ferromagnetic (FM), antiferromagnetic (AFM), and charge density wave (CDW) states, respectively. The tiny violet regions in (b) correspond to non-collinear and non-coplanar phases.}
\label{fig3}
\end{figure}

\begin{figure}[!t]
\hspace*{-0.5cm}
\vspace*{0cm}
\begin{overpic}[width=0.9\columnwidth]{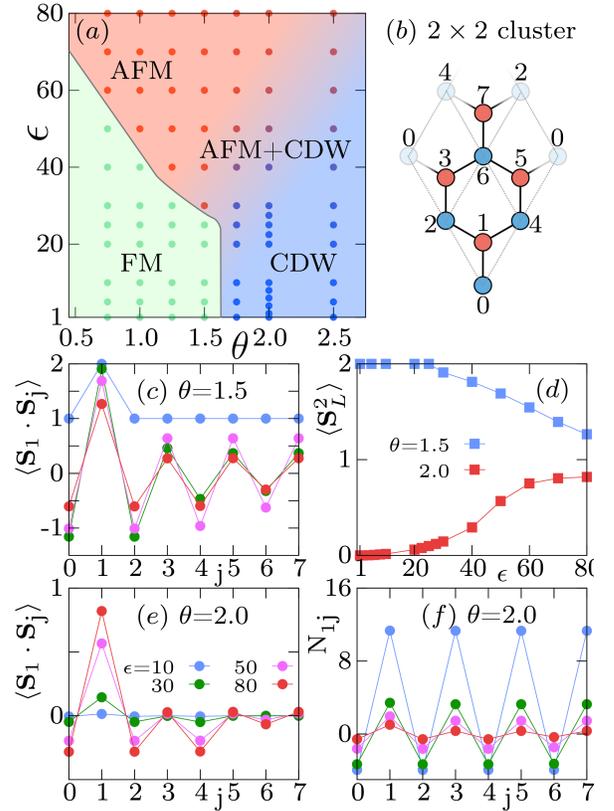}
\end{overpic}
\caption{(a) Effective dielectric constant $\epsilon$  vs twist angle $\theta$ phase diagram for the simplified KMH model by solving the small $2\times 2$ cluster using the Lanczos technique. The $2\times 2$ honeycomb cluster with periodic boundary conditions is shown in (b); the dashed thin lines depicts the underlying triangular Bravais lattice. (c,e) The spin-spin correlation with respect to site=1 ($\langle \mathbf{S}_{1}\cdot\mathbf{S}_{j} \rangle$) for various values of $\epsilon$, at fixed (c) $\theta=1.5$ and (e) $\theta=2.0$. (d) The local moment $\langle \mathbf{S}_{L}^2 \rangle$ for $\theta=1.5$ and 2.5 vs. $\epsilon$. (f) The density-density correlation with respect to site=1 ($\textrm{N}_{1j}$) for various values of $\epsilon$, at fixed $\theta=2.0$.
}
\label{fig4}
\end{figure}
{\it{Numerical results at half-filling.---}}
We create $\epsilon$ vs $\theta$ phase diagrams to investigate the physics of the KMH model at half-filling  $n$=$N/L$=$2$, where $N$ is the total number of fermions and $L$=$L_{1}$$\times$$L_{2}$ the total number of unit cells. We studied 6$\times$6 and 12$\times$12 system sizes using the unrestricted Hartree-Fock technique. We choose a broad range of $\epsilon$ $\in[1,80]$ as it can be tuned by changing the distance with the nearby metallic gate. Moreover, $\epsilon$ will be enhanced by the charge-fluctuations between the moir\'e bands considered here and other remote moir\'e bands. The $\epsilon$ vs $\theta$ phase diagram for the KMH model is shown in Fig.~\ref{fig3}(a). Surprisingly, in addition to the expected antiferromagnetic (AFM) state, we have unveiled two new states not anticipated to be stable: the S=1 ferromagnetic (FM) state for $\theta<1.75$ and the charge density wave (CDW) state for $\theta\ge1.75$. The non-local density-density repulsion plays the key role to stabilize the CDW state. The competition between the non-local FM direct exchange ($\propto \epsilon^{-1}(J_{02}+J_{13}+2J_{03})$) and the AFM superexchange ($\propto (\epsilon t^{2})/(U_{0}+J_{01})$) leads to the transition from FM to AFM state as $\epsilon$ increases. We found that the AFM state is present only for $\epsilon>20$ with local moment $\mathrm{S}<1$. See Fig.~\ref{fig3}(c,d,e) for the pictorial represention of the FM, AFM, and CDW states.

We also used the simplified KMH model, only keeping the first 4 terms in Eq.~\ref{KMH}, and found all three phases are present nearly in the same region of the phase diagram (see Fig.~\ref{fig3}(b)), suggesting that the FM direct exchange and the density-density repulsion are the most important non-local interactions for the half-filled KMH model.

To investigate the effect of the quantum fluctuations, we used the Lanczos technique and studied a small $2\times2$ cluster with periodic boundary conditions (Fig.~\ref{fig4}(b)), using the simplified KMH model. The phase diagram is shown in Fig.~\ref{fig4}(a). We again found the FM, AFM, and CDW states in the same region of the phase diagram. Fig.~\ref{fig4}(c) shows the spin-spin correlation, with respect to site=1, $\langle{\bf S}_{1}\cdot {\bf S}_{j}\rangle$ for $\theta$=$1.5^\circ$, depicting strong FM correlations for $\epsilon$=10, and AFM correlations for $\epsilon \in \{30,50,80\}$. Fig.~\ref{fig4}(e) and Fig.~\ref{fig4}(f) show $\langle{\bf S}_{1}\cdot {\bf S}_{j}\rangle$ and density-density correlations $N_{1j}=\langle n_{1}n_{j}\rangle-\langle n_{1} \rangle\langle n_{j}\rangle$, respectively, at fixed $\theta$=2.0 depicting the growth of AFM correlations and suppression in the CDW as $\epsilon$ increases. This indicates a smooth transition from the CDW phase to the AFM phase. We believe larger systems are required to confirm whether it is a 2nd-order phase transition or a crossover. The FM to AFM or the FM to CDW are first order transitions because the total spin suddenly changes from $2L$=$8$ to $0$. The averaged local moment ${\bf S}^{2}_{L}=(1/4L)\sum_{i\alpha}\langle {\bf S}_{i\alpha}^{2} \rangle$ as a function of $\epsilon$ is shown in Fig.~\ref{fig4}(d). We found, for $\theta<1.75$ that ${\bf S}^{2}_{L}$ decreases as $\epsilon$ is increased while the system transits from the S=1 FM to AFM phases, whereas for $\theta\ge1.75$, ${\bf S}^{2}_{L}$ grows with $\epsilon$ developing AFM correlations with weak CDW.

{\it{Conclusions.---}}
We showed that the twisted $\Gamma$-valley TMD bilayers contains physics beyond the conventional multi-orbital Hubbard model. We provide a KMH model which can be used to theoretically study the multi-orbital physics of TMD bilayers. Using our numerical studies at half-filled moir\'e bands we show that the non-local direct-exchange terms and density-density interactions can lead to S=1 FM insulators and CDW states, respectively, depending on $\epsilon$ and $\theta$. The AFM state can also be obtained but at large $\epsilon>20$. Our theoretical prediction of a S=1 FM insulator can be verified by measuring the magnetic susceptibility and Weiss constant in real materials~\cite{Tang01,YTang01}, and the charge ordered state can be observed using high-resolution scanning tunneling experiments~\cite{HLi01}. The KMH model can also be used for further theoretical investigations like doping near half-filled correlated insulators and for studying Mott-Wigner crystals at fractional fillings by including longer range density-density interactions. 

N. Kaushal and E. Dagotto were supported by the US Department of Energy, Office of Science, Basic Energy Sciences, Materials Science and Engineering Division. 






\end{document}